# BANKRUPTCY ANALYSIS USING IMAGES AND CONVOLUTIONAL NEURAL NETWORKS (CNN)

Luiz Wanderley Tavares[1]
José Afonso Mazzon[2]
Francisco Carlos Paletta[3]
Fábio Meletti de Oliveira Barros[4]

## ABSTRACT

The marketing departments of financial institutions strive to craft products and services that cater to the diverse needs of businesses of all sizes. However, it is evident upon analysis that larger corporations often receive a more substantial portion of available funds. This disparity arises from the relative ease of assessing the risk of default and bankruptcy in these more prominent companies. Historically, risk analysis studies have focused on data from publicly traded or stock exchange-listed companies, leaving a gap in knowledge about small and medium-sized enterprises (SMEs). Addressing this gap, this study introduces a method for evaluating SMEs by generating images for processing via a convolutional neural network (CNN). To this end, more than 10,000 images, one for each company in the sample, were created to identify scenarios in which the CNN can operate with higher assertiveness and reduced training error probability. The findings demonstrate a significant predictive capacity, achieving 97.8% accuracy, when a substantial number of images are utilized. Moreover, the image creation method paves the way for potential applications of this technique in various sectors and for different analytical purposes.

[1] Prof. Dr. Luiz Wanderley Tavares – lwt@usp.br – Orcid: 0000-0001-5013-8495
[2] Prof. Dr. José Afonso Mazzon - jamazzon@usp.br – Orcid: 0000-0003-1556-520X
[3] Prof. Dr. Francisco Carlos Paletta - fcpaletta@usp.br – Orcid: 0000-0002-4112-5198
[4] Prof. M.Sc. Fábio Meletti de Oliveira Barros - fabio.meletti@me.com – Orcid: 0009-0000-4084-8725

The address for all is:
Faculdade de Economia, Administração, Contabilidade e Atuária da Universidade de São Paulo.
Avenida Professor Luciano Gualberto, 908 - Butantã - São Paulo/SP - 05508-010 – Brazil.

1. INTRODUCTION

The evolving landscape of the financial sector has witnessed increased fragmentation, prompting scrutiny from both industry stakeholders and policymakers. This fragmentation exerts significant impacts on service efficiency, transparency, and global financial stability, as highlighted by Claessens (2019). Recognizing the imperative of fostering financial inclusion, particularly through digital channels, the World Bank and G20 have prioritized initiatives to stimulate economic development. However, the proliferation of digital banks and Fintech firms has raised concerns regarding the resilience of the economy, as noted by Ozili (2018).

Amidst this backdrop, distinctions arise between Fintech firms, which cater to niche markets under adaptable regulatory frameworks, and Digital Banks, which adhere to traditional banking regulations, as articulated by Wewege and Thomsett (2020). In response to shifting consumer preferences, banking institutions are recalibrating their marketing strategies towards cultivating customer loyalty, as observed by Verona (2004). In light of these developments, this study advocates for the utilization of accounting data as a means to assess the financial health of small and medium-sized enterprises (SMEs), aiming to provide nuanced insights beyond conventional approaches.

2. CONTEXTUALIZATION

Recent developments in the financial sector have engendered a paradigm shift, characterized by escalating fragmentation. This phenomenon has garnered increased scrutiny from industry stakeholders and the polity alike. Fragmentation's manifestation varies across different financial domains and can be attributed to an array of factors transcending financial regulations and oversight mechanisms. The implications of fragmentation are multifaceted, affecting the efficiency of financial service delivery, impinging upon transparency, and impacting consumer and investor safeguards; it also bears significance for global financial equilibria (Claessens, 2019).

Driven by the World Bank and the G20 since 2010, the impetus towards financial inclusion has been strategically underscored. Aiming to attenuate socio-economic disparities and spur global economic development, the World Bank has championed digital financial inclusion as a cornerstone for the advancement of nations with burgeoning economies.

This initiative of digital financial inclusion has materialized through the establishment of digital banking entities and Fintech firms, which has catalyzed economic expansion via an upsurge in financial transactions within the system. However, this growth raises questions about the potential exacerbation of economic downturns under market duress (Ozili, 2018).

Digital Banks and Fintechs share a foundational premise as financial institutions predominantly operational through digital infrastructures, obviating the necessity for physical branches. Despite this similarity, Digital Banks largely mirror traditional banking services and adhere to the regulatory and operational stipulations of central banks. Conversely, Fintechs innovate by tailoring services to niche markets, underpinned by more flexible regulatory frameworks and operational prerequisites.

These entities leverage a competitive edge through a user-centric mobile experience. Traditional banks, in response, are expediting their digital metamorphosis, integrating innovative partnerships with Fintechs to enhance customer experience, thereby achieving cost-effectiveness, security, and convenience (Wewege & Thomsett, 2020).

Wewege and Thomsett (2020) posit that Fintechs will perpetuate the trajectory of technological advancement and agile implementation, with a concerted focus on user experience and customer value proposition through expedited process innovation.

Historically, the marketing strategies of banks were circumscribed to a simple array of incentives aimed at customer acquisition. Subsequent insights led to the realization of the necessity for cogent marketing planning and oversight. A bank's expertise in advertising and innovation must be underpinned by strategic planning and control mechanisms to accurately gauge market potential, orchestrate objectives, and measure outcomes to avoid significant performance deficits. The pivot towards relationship-building and customer fidelity has emerged as a prevailing trend in bank marketing (Verona, 2004).

Notwithstanding these advancements, both traditional and digital-first banking institutions continue to refine their marketing endeavors and delineate the risk profiles intrinsic to their fiscal activities.

This study unveils an innovative approach to company risk assessment, leveraging data derived directly from accounting records. This method enables the scrutiny of small and medium-sized enterprises (SMEs) where accounting practices are typically simplified. Hence, financial institutions often resort to superficial analyses based on gross revenue, overlooking deeper financial insights.

2.1. First studies on corporate bankruptcy

Financial risk analysis critically assesses the propensity of clients to fulfill their payment obligations under negotiated financial terms. In a seminal study, Altman (1968) sought to devise a predictive model of corporate fiscal performance. Prior to Altman's work, Myers (1963) had pioneered efforts to derive a quantifiable measure of creditworthiness for individuals. These early endeavors, including Myers' and those that succeeded in the pre-machine learning era, utilized demographic and socioeconomic data variables, encompassing factors such as gender, number of dependents, marital status, duration of residence, occupational role, and monthly income, among others.

Altman (1968) innovated in this domain by analyzing the annual financial statements of corporations and deriving a set of financial ratios from these records. He initiated his research with a cohort of sixty-six enterprises across diverse sectors, subsequently filtering out smaller entities to focus exclusively on five pertinent financial ratios.

The pursuit of accessible corporate financial data has historically been a challenge. Publicly available data is typically sourced from publicly traded firms, which are mandated to publish their financial outcomes. This scenario provides, in the best cases, access to a span of three to five years of quarterly financial data, thereby assembling a qualified database for analysis contingent on the dataset's volume.

Large enterprises, irrespective of their public trading status, have been traditionally subject to analysis premised upon their financial statements. Conversely, banks possess limited informational insight into small and medium-sized enterprises (SMEs), leading to the provision of standard financial products that may not be congruent with the SMEs' authentic financial conditions.

## 2.2. Recent studies using machine learning

Altman's foundational work in 1968 operationalized discriminant analysis as the cornerstone for the assessment of financial health. Building upon this, Frydman et al. (1985) leveraged recursive partitioning algorithms to examine corporate financial distress, juxtaposing their approach with Altman's discriminant analysis. The deployment of decision trees facilitated a nuanced analysis of the selected financial indices.

Shumway (2001) employed Altman's study as a comparative framework for his bankruptcy prediction model, yet it rendered an accuracy of less than 70%. Kassai and Onusic (2004) adopted data envelopment analysis for insolvency prediction, using general and long-term debt ratios, along with debt composition as inputs for their model. They reported an accuracy of 90.0% for insolvent companies and 74.0% for solvent ones, cumulatively achieving a classification accuracy of 76.6%. It is noteworthy, however, that the sample comprised a modest total of 60 companies, among which ten were insolvent.

In his 2006 study, Carton and Hofer innovated with a predictive model grounded in regression analysis of financial ratios categorized into six metrics: profitability, growth, efficiency, cash flow, survivability, and market valuation. Confirmatory factor analysis was employed to substantiate the categorization, which led to the model attaining an accuracy rate slightly above 80%. The study's significant contribution lay in the stratification of financial indices and validating this categorization through factor analysis.

Min et al. (2006) introduced a hybrid model integrating genetic algorithms with support vector machines (SVM), designed to enhance SVM's predictive efficacy through feature subset selection and parameter optimization. The model demonstrated improved accuracy in bankruptcy prediction, attaining an 80.3% validation accuracy.

The implications of macroeconomic conditions on credit risk models were scrutinized by Carling et al. (2007), who applied Monte Carlo simulations to assess value at risk and projected deficits, correlating the likelihood of default with both company-specific and macroeconomic factors.

Further, Hung and Chen (2009) conducted comparative analysis using decision trees, backpropagation neural networks (BNN), and SVM. Utilizing a dataset encompassing 30 financial indices from 120 companies, their findings indicated that the BNN method achieved a peak accuracy of 72.5%.

In a comparative analysis of machine learning techniques, Aktan (2011) examined the efficacy of eight machine learning algorithms—Naive Bayes, Bayesian Network, k-NN, ANN, SVM, C4.5, CHAID, and CART—in the context of financial distress. Utilizing a dataset of fifty-three financial indices, the study identified the CART decision tree as yielding the highest accuracy. Notably, the Naive Bayes and Bayesian Network algorithms also demonstrated superior performance relative to the ANN.

Barboza, Kimura, and Altman (2017) built upon Carton's (2004) categorical framework within machine learning models, introducing additional financial categories such as operating margin and change in return on equity to refine bankruptcy prediction. In their empirical analysis, the random forest algorithm emerged as the most accurate method, achieving a remarkable 87.1% accuracy. This study also evidenced the efficacy of bagging and boosting techniques in comparison to Carton's earlier model.

Addo et al. (2018) conducted an extensive comparison of seven predictive models, including an elastic network (enhanced logistic regression), random forest, gradient boosting, and neural networks of varying complexities. The initial phase utilized 181 variables across a dataset exceeding 110,000 financial contracts, subsequently focusing on the 10 most pivotal variables as determined by the models. The findings indicated that the neural network models did not surpass the random forest and gradient boosting models, which both achieved accuracies above 97%, with some instances exceeding 99%. However, the pre-evaluated nature of the companies in the concluded contracts introduced a selection bias, discernible from the fact that only 1.5% of the records pertained to defaulted contracts, potentially skewing the results.

Hosaka (2019) implemented convolutional neural networks (CNNs) to forecast bankruptcies, harnessing images that depicted financial ratios. This study elucidated the aptitude of CNNs for image analysis over traditional numerical data processing. Employing a compendium of 7,520 images, the CNN, based on GoogLeNet, was trained to discern between bankrupt and ongoing firms, with the predictive accuracy surpassing that of decision trees, linear discriminant analysis, support vector machines, and multilayer perceptrons. It is noteworthy, though, that the method was not tailored to investigate the root causes of bankruptcy, and the study did not delineate the rationale for selecting a GoogLeNet-based CNN architecture.

In an exploratory study, Bussmann et al. (2021) utilized a supervised extreme gradient boosting machine learning model, contrasting it with logistic regression to examine potential defaults in small to medium-sized enterprises (SMEs). The selected sample comprised 15,000 such enterprises, each tagged with a Boolean variable indicating default, along with a suite of conventional financial and accounting indices. These entities were included in the dataset by virtue of having applied for loans. The dataset was partitioned into a training subset (80%) and a testing subset (20%), with the extreme gradient boosting method achieving a notable accuracy of 93%. This important level of accuracy could be attributed to the fact that all enterprises in the dataset had existing financial obligations; notably, the model was not evaluated on non-debtor companies. The authors advocate for subsequent analyses to employ imbalanced datasets, where default instances are infrequent, positing that the scarcity of such events could significantly alter the predictive accuracy of the model.

Regarding time-variant data, time series analysis is traditionally employed. However, recent investigations have incorporated neural network methodologies, including convolutional neural networks (CNNs) as noted by Jin et al. (2020), recurrent neural networks (RNNs) as explored by Madan and Mangipudi (2018), and artificial neural networks (ANNs) with multilayer perceptrons (MLPs), as utilized by Aktan (2011), all reporting efficacious outcomes.

Aktan's (2011) research specifically highlighted a decrement in model accuracy correlated with the age of data: 90.0% for one-year-old data, 85.7% for two-year-old data, and 76.3% for three-year-old data. Such deterioration is indicative of a well-documented phenomenon within temporal models known as the "recency effect" or "recency bias," wherein more recent data disproportionately influences the model's predictions.

Zhao et al. (2017) elucidated the efficacy of convolutional neural networks (CNNs) in processing time-series data. Hosaka (2019) further underscored the aptness of CNNs for image applications, employing them in an analysis of grayscale images that encapsulated the financial indices derived from annual balance sheets. This approach yielded promising results, although it was noted that extensive image datasets were required to effectively train the CNN model, posing a constraint on the utility of CNNs.

Prevailing research, including that by Jin et al. (2020) and Hosaka (2019), has encountered limitations, notably the reliance on limited sample sizes and the use of data sourced from public companies within singular national contexts. Moreover, the frequency of data collection—often on an annual or, less commonly, quarterly basis—also limits the temporal resolution of the analysis.

The present study proposes a novel methodology for assessing the fiscal health of small and medium-sized enterprises (SMEs) using comprehensive financial data extracted directly from accounting systems. This approach seeks to predict potential defaults or bankruptcies by analyzing recent financial data pertaining to SMEs.

Leveraging complete accounting records, this methodology facilitates the creation of a detailed time series from SME balance sheets. These balance sheets, consisting of monthly accounting entries, enable the monitoring of variations across individual accounts and the computation of financial ratios monthly, thereby enhancing the granularity of financial health assessments for SMEs.

Post-2010, the utilization of machine learning for predictive modeling with annual data has yielded notable success. Nonetheless, balance sheets represent only temporal financial snapshots, lacking dynamic progression.

Innovatively, the assembly of monthly balance sheets can conjure a series of 'financial photographs,' collectively forming an 'animation' that depicts the financial evolution of a company over time. While conventional time series methods can be applied to such data, they necessitate a degree of preprocessing for optimal utility.

Recent developments have seen the integration of neural networks capable of processing time series data. These networks excel particularly in image processing, which aligns with the objectives of this research—to construct a model capable of predicting default or bankruptcy for small and medium-sized enterprises (SMEs) by interpreting their financial data as processed images. By feeding such images into a neural network, it is envisioned that the system can prognosticate potential fiscal distress. Consequently, this model endeavors to encapsulate the financial trajectory of an SME within an image, thus enabling the neural network to discern the financial state of the company.

## 3. DATA SOURCE

This study accessed archival data from 580 accounting firms, encompassing comprehensive financial records of approximately 105,000 small and medium-sized enterprises (SMEs) across various sectors of the Brazilian economy. This extensive dataset included over 20 million records of accounting charts and more than 400 million accounting entries spanning the last decade.

The 400 million accounting entries, documented daily, were systematically organized into monthly financial accounts, forming a historical time series for the cohort of 105,000 SMEs.

Data diversity was marked by the utilization of 15 disparate accounting systems, each with its unique data structuring methodology. Despite accounting standards prescribing uniform nomenclature for financial accounts, discrepancies are commonplace, with accountants employing varying codes and systems to denote identical financial transactions. Such variation often precludes direct data comparability across different enterprises.

This lack of standardization poses significant analytical challenges, hindering the direct comparison of one company's financial data with another's. Historically, researchers have relied on consolidated data to circumvent these issues, benefiting from its uniformity. Nonetheless, consolidation tends to obscure granular details, which could otherwise yield insights into an enterprise's financial status and contribute to a richer comparative analysis.

To address this, the preliminary phase of the research involved the normalization of financial data, synchronizing the accounting records across all firms to a unified chart of accounts. This study synthesized the disparate accounting information into 208 standardized financial accounts.

Subsequently, a comprehensive monthly balance sheet for each SME was compiled, forming a vector representation that encompassed all accounting data per enterprise, monthly.

Prior to deploying machine learning techniques, financial indices and their computation traces were aggregated into a singular record for each enterprise per month. Additionally, indices reflecting month-to-month data variations were formulated.

To mitigate inaccuracies arising from the analysis of absolute financial figures, all such data were normalized by scaling them to percentages relative to the size of the firm, thereby facilitating direct comparisons across SMEs of varying scales. This normalization was achieved by expressing all values as proportions of the total asset value of each company.

## 4. METHODOLOGY

The primary objective of this research was the standardization of accounting data across companies to facilitate a consistent analytical framework. The normalization process was designed to generate a structured array, constituting a 'month-by-accounting' vector for each firm, encapsulating the comprehensive monthly accounting activities.

Upon constructing these vectors, a subsequent transformation was conducted to transmute the numerical vectors into visual representations. In these resultant images, the temporal dimension was mapped along the vertical axis, displaying the months, while the horizontal axis corresponded to the selected accounting accounts. This innovative approach allows for the multidimensional financial data to be analyzed through visual means, potentially revealing patterns and trends that may not be as apparent in traditional tabular data formats.

### 4.1. Normalization of the chart of accounts

The methodology commences with the harmonization of the Charts of Accounts and the organization of associated accounting entries. This preliminary step is crucial to ensure data uniformity, thereby enabling comparative analysis.

Gans et al. (2007) elucidate that the Chart of Accounts constitutes the foundational framework of accounting information systems within various organizations and industries. These charts are essentially compendiums of account titles and their associated numerical codifications, which are instrumental in systematically recording financial transactions, including revenues and expenses, and categorizing assets and liabilities. By adopting a Chart of Accounts, financial information can be methodically classified and compiled into functional aggregates such as product lines, cost centers, and operational departments, tailored to meet the discrete requisites of an organization or industry. These accounts are pivotal in generating uniform financial statements and managerial reports that are indispensable to both internal management and external stakeholders. The Chart of Accounts is, therefore, an integral element of both public and private sector accounting information systems.

On the global front, the International Financial Reporting Standards (IFRS), promulgated by the International Accounting Standards Board (IASB), aim to bring coherence to the financial reporting of enterprises within the financial sphere. Nonetheless, Trimble (2017) posits that the practical application of IFRS is fraught with challenges. Often, the dichotomy between local accounting practices and the international norms established by IFRS leads to variances that hinder the full-fledged adoption of these standards.

In the practical application of accounting principles, a plethora of analogous charts of accounts is evident across companies. Nonetheless, discrepancies within these charts preclude the possibility of direct comparison of account codes between different companies' financial records.

To address this issue, the study established a standardized chart of accounts comprised of four hierarchical levels, encompassing a total of 208 individual accounts.

This structure included 4 accounts at level one, 12 at level two, 42 at level three, and 150 at level four. The highest-level accounts typically represent broad categories such as "Assets," "Liabilities," "Revenues, Costs, and Expenses," and "Contra Accounts." These broader categories can be aligned and compared using straightforward methodologies. However, the level four accounts present a greater challenge due to their complexity; for instance, an account labeled "Customer" may appear under both "Current Assets" and "Non-Current Assets," leading to potential ambiguities.

Despite these complexities, a pattern of comparability emerges when descriptions at all levels are concatenated. These comparative patterns are noticeable in Table 1, which allows for a more granular and detailed examination of account configurations.

| Charts of Accounts with 4 Levels of Detail | | | | | Full Account Description |
|---|---|---|---|---|---|
| Accounting Account | | | | Account Description | |
| 1 | | | | Asset | Asset |
| 1 | 1 | | | Current assets | Asset Current assets |
| 1 | 1 | 1 | | Cash and cash equivalents | Asset Current assets Cash and cash equivalents |
| 1 | 1 | 1 | 001 | Cash | Asset Current assets Cash and cash equivalents Cash |
| 1 | 1 | 1 | 002 | Banks | Asset Current assets Cash and cash equivalents Banks |
| 1 | 1 | 2 | | Accounts receivable | Asset Current assets Accounts receivable |
| 1 | 1 | 2 | 001 | Customers | Asset Current assets Accounts receivable Customers |
| 1 | 1 | 2 | 002 | Duplicates | Asset Current assets Accounts receivable Duplicates |

Table 1: Concatenation patterns for descriptions of charts of accounts.

Within Table 1, the methodical concatenation of account descriptions yields sentences that are capable of comparison. For instance, phrases such as "Current Assets Accounts Receivable from Customers" or "Current Assets Accounts Receivable Duplicates" display a substantial semantic proximity to "Current Assets Accounts Receivable." This degree of relational closeness is not maintained when these phrases are contrasted with a dissimilar description, such as "Active Non-Current Cash and Cash Equivalents."

The crux of the challenge lies in condensing a vast array of accounts into a succinct chart that can uniformly accommodate all variations of customer-related accounts. The envisioned transition from the original to the newly assigned codes would be facilitated through machine learning algorithms, trained on the comprehensive descriptions aligned with the new account codes. The primary objective of this compact chart of accounts is to facilitate the calculation of financial ratios, enabling month-to-month comparative analysis of a company's financial trajectory against its industry counterparts.

The development of these 208 accounts was informed by this requirement and mirrors the contemporary fiscal status of the entities within the research database. Although the final chart consisted of 208 accounts, the training process incorporated 338 complete descriptions, indicating that multiple descriptors could correspond to a single account. This approach ensures that the model accounts for the diverse notations employed by different accounting professionals.

The encoding process was implemented using the Universal Sentence Encoder, developed by Cer et al. (2018), which is designed to transform sentences into high-dimensional embedding vectors. This tool is leveraged for transfer learning applications in various Natural Language Processing (NLP) tasks. Its design is conducive to enhancing the performance of language processing tasks, which include, but are not limited to, assessing semantic similarity, facilitating text classification, and enabling clustering algorithms.

Consequently, distinct vectors were generated for each of the 338 comprehensive descriptions associated with the 208 designated target accounts. This methodology allows for the swift conversion of a diverse accounting chart into the standardized 208-account framework by aligning each description with its nearest complete equivalent. Subsequently, the appropriate standard account is integrated into the comprehensive financial transaction system.

The repository of full descriptions utilized for training can be expanded to refine the accuracy rates observed in the accounting accounts of the examined charts. The current training regimen has been validated to ensure a minimum accuracy threshold of 86.5%.

4.2 Generation of the accounting vector

Following the standardization of cost centers, vectors encapsulating the monthly financial data were constructed. These vectors not only encompass the financial data but also detail the operational sector and geographic location (state and country) of each enterprise.

This dataset was compiled by aggregating all financial transactions for each company monthly, thereby synthesizing a monthly balance sheet for every month the company has been operational. After generating the trial balance, a comprehensive calculation of financial ratios for the month was conducted, alongside the implementation of horizontal and vertical analyses for the financial accounts.

The selection of financial indices for inclusion in the predictive model vectors is delineated in Table 2. This list comprises twenty-one financial indices that are conventionally employed in market analyses.

These indices are derived from specific accounting entries, which may be referenced multiple times across different indices. This recurrence poses a risk of introducing bias into the data when subjected to statistical or machine learning algorithms.

In conventional practice, the accounts constituting the trial balance undergo both horizontal and vertical analysis. Horizontal analysis critically examines the fluctuations in account figures over successive periods, identifying trends and patterns in the data.

| Selected Financial Ratios | |
|---|---|
| Quick ratio | |
| Immediate Liquidity | |
| Permanent Assets | % Permanent Assets |
| Liquidity Ratio | |
| Fixed Resources | |
| Fixed Assets | |
| Debt Ratio | |
| Net Working Capital | % Net Working Capital |
| Own Working Capital | % Own Working Capital |
| Long Term Assets | % Long Term Assets |
| General Liquidity | |
| EBITDA | % EBITDA |
| Operating Expenses | % Operating Expenses |
| Financial Expenses | % Financial Expenses |
| Net Margin | |
| Contribution Margin | |
| Tax Burden | |
| Gross Margin | |
| Expenses | % Expenses |
| Growth | |
| Balance | |

Table 2: Selected Financial Ratios.

Vertical analysis complements this approach by quantifying each financial account as a proportion of the total assets held by the enterprise. This analytical technique is instrumental in evaluating financial performance over time, providing insights into a company's operational efficacy, financial structure, and comparative standing within the industry.

Consequently, a vector was composed of the accounting accounts that are most frequently implicated in financial indices and that possess a significant capacity to reflect the fiscal health of a business. Calculations were performed to determine the relative contributions of each account, as delineated by horizontal and vertical analyses. The accounts incorporated into this vector, which were carefully chosen for their diagnostic value, are enumerated in Table 3.

| Standardized Accounts | |
|---|---|
| Number | Description |
| 11100 | Current assets available in cash |
| 11200 | Current assets customers |
| 11300 | Current assets stock |
| 11500 | Current assets financial investments |
| 13200 | Non-current assets investments |
| 13300 | Fixed non-current assets |
| 13400 | Intangible non-current asset |
| 21100 | Current liabilities suppliers |
| 21200 | Current liabilities loans and financing |
| 22000 | Non-current liabilities |
| 22300 | Long-term non-current liabilities |
| 23000 | Liabilities net worth |
| 31100 | Sales revenue |
| 32100 | Costs of products, goods and services sold |
| 32200 | Operational expenses |
| 32300 | Financial expenses |
| 32400 | Other operating expenses |
| 32500 | Provision for corporate income tax and social contribution on net profit |
| 32700 | Direct costs of producing services |
| 32800 | Costs of services provided |
| 32900 | Operating expenses administrative expenses |

Table 3: Accounting accounts selected

4.3 Sampling

In Brazil, the categorization of economic activities and the corresponding classification criteria are orchestrated by various tax administration entities through the CNAE framework (National Classification of Economic Activities). The CNAE system stratifies businesses into distinct levels such as divisions, groups, classes, and subclasses, allowing for detailed economic segmentation.

To validating the methodology, three specific divisions were selected: 'retail trade,' 'human health care activities,' and 'office services, administrative support, and other business-oriented services.' Within the broader dataset of 105,000 enterprises, a subset of 14,456 companies was extracted based on these categories.

The analysis was further refined by focusing on two retail trade groups, identified as 472 and 475. The selection of these groups, as detailed in Table 4, was informed by their representation of a substantial portion of the commercial demographic within the dataset.

| Retail business | | | |
|---|---|---|---|
| Division | Group | Companies | Description |
| 47 | 471 | 1.051 | Non-specialized retail trade |
| 47 | 472 | 1.131 | Retail trade of food products. drinks and smoke |
| 47 | 473 | 389 | Retail trade of fuels for motor vehicles |
| 47 | 474 | 1.266 | Retail trade of construction materials |
| 47 | 475 | 1.545 | Retail trade of computer and communication equipment. household equipment and articles |
| 47 | 476 | 377 | Retail trade of cultural items. recreational and sports |
| 47 | 477 | 1.111 | Retail trade of pharmaceutical products. perfumery and cosmetics and medical articles. optical and orthopedic |
| 47 | 478 | 2.659 | Retail trade of new products not previously specified and used products |
| Total 47 | | 9.529 | |

Table 4: Retail trade with the number of companies.

The quantification of firms engaged in human health care activities is systematically cataloged in Table 5. This division was selected to contrast with the initial sector of focus due to its distinct nature. Additionally, the presence of a substantial quantity of companies within this sector in the database provided a robust basis for analysis.

| Human Health Care Activities | | | |
|---|---|---|---|
| Division | Group | Companies | Description |
| 86 | 861 | 241 | Hospital care activities |
| 86 | 862 | 6 | Mobile emergency care and patient removal services |
| 86 | 863 | 1.746 | Outpatient care activities performed by doctors and dentists |
| 86 | 864 | 257 | Diagnostic and therapeutic complementary service activities |
| 86 | 865 | 582 | Activities of health professionals. except doctors and dentists |
| 86 | 866 | 31 | Health management support activities |
| 86 | 869 | 56 | Human health care activities not previously specified |
| Total 86 | | 2.919 | |

Table 5: Human health care activities with the number of companies.

The composition and codes of Division 82 are detailed in Table 6. This division was specifically chosen to function as a validation set for the modeling approach, enabling the testing and refinement of the training process across another distinct sector.

| Office Services. Administrative Support and Other Services Provided Mainly to Companies | | | |
|---|---|---|---|
| Division | Group | Companies | Description |
| 82 | 821 | 1.373 | Hospital care activities |
| 82 | 822 | 16 | Mobile emergency care and patient removal services |
| 82 | 823 | 216 | Outpatient care activities performed by doctors and dentists |
| 82 | 829 | 403 | Human health care activities not previously specified |
| Total 82 | | 2.008 | |

Table 6: Office, administrative support, and other services provided to companies

Regarding the dataset employed for the analysis, a total of 416,710 monthly data vectors were compiled for Division 47, with an additional 116,487 vectors for Division 86, and 75,373 for Division 82. These comprehensive datasets were systematically integrated into a high-performance SQL database to ensure efficient data handling and retrieval for subsequent analyses.

5. MODEL DEVELOPMENT

The aim of this research is to develop a predictive data model to discern which companies may be on the brink of default or insolvency, an analytic tool of considerable utility for the financial sector in managing risk and tailoring customer-specific financial products.

To facilitate this, the dataset encompassing 14,456 enterprises was dichotomously labeled, with '1' signifying companies that have defaulted or gone bankrupt, and '0' denoting active and operational companies. Public entities demonstrating inactivity were also categorized under '1'. This binary classification was consistently applied across all instances within the model training process.

Traditionally, company performance assessments via balance sheets can be conducted either through derived financial ratios or through horizontal and vertical analyses. Balance sheets, while reflective of a company's financial status at a discrete point in time, are less effective at conveying the financial dynamics that transpired in preceding periods (Krugman, 1999), thereby limiting the fidelity of year-over-year comparative analyses due to a lack of intermediate temporal data. However, the corpus of this study is enriched with monthly data, thus offering a more granular temporal perspective of company performance.

The primary challenge was to encapsulate the temporal trajectory of a company's financial history over the 12 months leading up to the final balance sheet or cessation of operations. Leveraging neural networks' proficiency in image processing, as opposed to numerical data analysis, financial variables from the last 12 months of the 14,456 companies were transformed into an image format. In these images, each pixel corresponds to a financial variable for a given month, with the pixel's hue and intensity representing variations.

CNNs, renowned for their superior image analysis capabilities, were thus posited as the optimal methodology for examining such image-based representations of financial time series. An exemplar of such an image, encapsulating twelve months of accounting data for a single enterprise, is presented in Figure 1.

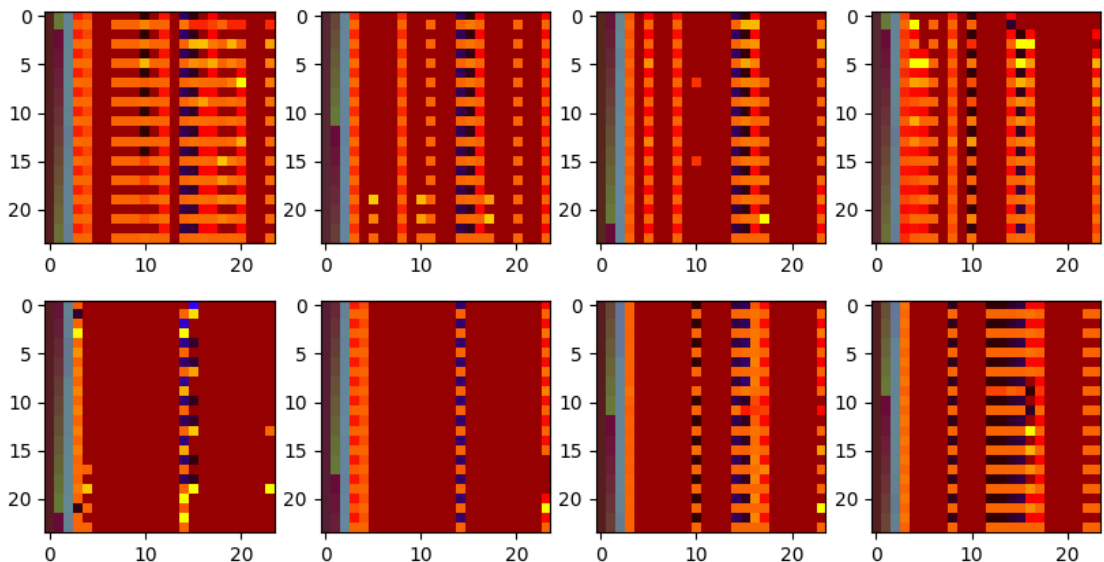

Figure 1: Twelve-month images of the accounting accounts of eight companies.

The imagery, derived from the financial data, is designed to be processed by a convolutional neural network (CNN), which is tasked with the objective of discerning patterns indicative of potential fiscal distress in companies. To implement the deep learning architecture for this task, the Keras framework was selected for its user-friendly interface and integration with Python, a programming language with widespread adoption in the computational research community (Chollet, 2021).

5.1. Image creation

The synthesized images, structured in a 24x24 grid, utilize the RGB color model, where each color is depicted by a triplet of integers ranging from 0 to 255. This numerical range quantitatively represents the intensity of red, green, and blue—hence the acronym RGB. A vivid example would be the representation of pure blue, which is coded as (0, 0, 255).

The first pixel in each image encodes the company's division and group, as well as the geographical state. Specifically, the first number denotes the division, the second the group (the group number multiplied by 10), and the third corresponds to the state code. For instance, a firm situated in São Paulo within group 472 would be signified by the pixel (47, 20, 35).

Subsequently, the second pixel delineates the period and country, with the first value representing the year (offset from 1970 and doubled to fit within the 0 to 255 range), the second the month (multiplied by 10), and the third the International Telecommunication Union (ITU) country code. For example, a Brazilian company operating in January 2020 would be depicted by the pixel (100, 10, 55).

The third pixel captures inflation data, where the first value is fixed at 100 to indicate the index type (using IPCA—Broad National Consumer Price Index), the second is the square root of the monthly percentage change multiplied by 100, and the third reflects the 12-month variation using a similar computation. The baseline figure of 125 represents zero, with only the integer part of the resulting sum utilized. Therefore, an inflation scenario of 0.54% monthly and an annual rate of 7.02064% (June 1997's figure) translates to the pixel (100, 132, 151).

Opting for a 24x24 format enriches the image with a greater density of information, with two rows allocated to the period representation. In the financial ratios-based images, both lines for the period replicate identical data. Conversely, in the accounting accounts-based images, one row displays the vertical analysis, while the other reveals the horizontal analysis. As for the images encoding indices, although capable of depicting two years of data, this research focuses on a 12-month span to facilitate a comparative analysis between the two distinct image types generated.

5.2. Algorithmic Training

Initial training was conducted using data from Divisions 47 and 86, employing both the financial ratio and accounting account images. The partitioning of the data for this phase allocated 80% of the images for model training and validation, with the remaining 20% designated for testing. Within the training subset, a further distribution allocated 90% for actual model training and 10% for validation. This segregation of images for the purposes of training and testing was executed using a random selection process, facilitated by the Keras framework's inherent capabilities. The outcomes of this process were evaluated by documenting the model's loss and accuracy metrics during the training phase, as well as the accuracy of the confusion matrix for the test image set.

After the individual training of each division, a combined dataset incorporating both Divisions 47 and 86 was constructed, utilizing both sets of images—those illustrating financial ratios and those depicting accounting accounts.

To augment the robustness of the model, Division 82 was also trained separately and then integrated with the previously trained divisions. This approach aimed to consolidate the model's learning across varied datasets, thereby enhancing its predictive performance.

5.3. Results

The efficacy of the neural network training was evaluated by analyzing the model's loss and accuracy. This analysis provided insights into the performance and predictive reliability of the neural network throughout the training process. Additionally, the outcomes for the test images were meticulously assessed using Precision, Recall and F1 Score, which offered a detailed representation of the model's predictive capabilities, including its precision in classifying the data correctly.

| Training # | Image Type | Training | | | Test | | Loss | Accuracy | Precision | Recall | F1 Score |
|---|---|---|---|---|---|---|---|---|---|---|---|
| | | Division / Group | Images Training | Images Validation | Division / Group | Images | | | | | |
| 1 | Financial Ratios | 47 | 4.618 | 513 | 47 | 1.286 | 24,2% | 92,2% | 96,8% | 89,9% | 93,2% |
| 2 | Accounts | 47 | 4.661 | 517 | 47 | 1.239 | 8,5% | 97,5% | 99,4% | 95,6% | 97,4% |
| 3 | Financial Ratios | 86 | 1.475 | 163 | 86 | 405 | 77,2% | 58,9% | 68,1% | 87,4% | 76,5% |
| 4 | Accounts | 86 | 1.470 | 163 | 86 | 410 | 65,8% | 65,0% | 69,3% | 83,6% | 75,8% |
| 5 | Financial Ratios | 47 and 86 | 6.075 | 675 | 47 e 86 | 1.710 | 10,9% | 95,7% | 96,5% | 97,1% | 96,8% |
| 6 | Accounts | 47 and 86 | 6.057 | 673 | 47 e 86 | 1.730 | 10,4% | 96,4% | 99,1% | 96,8% | 97,9% |
| 7 | Accounts | 82 | 973 | 108 | 82 | 270 | 71,1% | 50,0% | 75,9% | 68,2% | 71,9% |
| 8 | Accounts | 47, 82 and 86 | 7.040 | 782 | 47, 82 and 86 | 1.989 | 11,2% | 96,3% | 99,1% | 93,7% | 96,3% |

Table 7: Consolidated data from the tests.

In the context of convolutional neural networks, the loss test functions as a critical measure of a model's efficacy and dependability. This test employs a mathematical function designed to quantify the divergence between the model's predictions and the actual or anticipated outcomes. The resulting metric, expressed as a numerical value, gauges the accuracy of the model by quantifying losses, with lower scores denoting heightened accuracy. According to Yu et al. (2020), loss values exceeding 0.20 may signal potential overfitting within the model. Therefore, it is imperative that the loss test be interpreted in tandem with the model's accuracy metrics.

The accuracy metric serves as a benchmark for model performance, assessing the ratio of correct predictions to the total predictions made, as delineated by LeCun et al. (2015). However, the Precision, Recall, and F1 Score indices enhance the model's analytical capabilities.

For the analysis of test images, the confusion matrix is a pivotal tool, offering a structured overview of the correctly and incorrectly classified instances. Given the binary nature of the classification in this study, the matrix delineates the counts of true positives, false positives, true negatives, and false negatives. As depicted in Table 7, the test's accuracy was ascertained by dividing the aggregate number of correct classifications, both positive and negative, by the total number of tests executed.

5.3.1. Influence of the number of images in the training phase

The empirical results demonstrate that the volume of images utilized during the training phase of the convolutional neural network (CNN) has a pronounced impact on the model's performance. Independent of the nature of the data generating the images, there is a clear trend indicating improvements in both accuracy and loss metrics with an increase in the quantity of training images.

This correlation between the dataset size and CNN performance is consistent with findings across numerous studies employing CNN methodologies. The most favorable outcomes within this research were achieved when the training set exceeded 6,500 images.

A noteworthy pattern was detected regarding model loss in scenarios where fewer than 1,100 images were used for training; however, establishing a threshold for the minimum number of images conducive to the least loss was not the primary objective of this evaluation.

Training iterations comprising no more than 1,480 images consistently resulted in losses exceeding 60% and a training accuracy not surpassing 65%. Conversely, when the image count was elevated beyond 4,610, the lowest recorded accuracy was 92.2%, although this did not uniformly translate to a low rate of model loss—as evidenced by a 24.2% loss rate in the third training iteration.

The quantity of images, therefore, emerges as a critical factor influencing the enhancement of CNN outcomes. Specifically, when analyzing accounting data, models trained with image datasets larger than 4,610 displayed a potent predictive capability.

While the requirement for extensive image datasets may constitute a limitation in certain applications of CNNs, especially for object or trend recognition, it is not a constraining factor within the scope of this study. The robust dataset at hand adequately demonstrates the CNN's proficiency in accurately classifying images that may be deemed abstract from a human visual perspective.

5.3.2. Comparison between the use of financial ratios and accounting accounts

In the case of Divisions 47 and 86, images were crafted to represent both financial ratios and accounting data—including both vertical and horizontal analyses—with the goal of discerning potential disparities in predictive outcomes between these two approaches.

Financial indices are composed of underlying accounting data, with certain data points contributing to multiple indices. The inquiry at hand sought to ascertain if any correlations among these indices might affect the accuracy of predictions when compared to using raw accounting data.

When examining the data from Division 47, an assessment of variable correlations was conducted. This examination revealed a low correlation among most variables across the three distinct representations—financial ratios, and vertical and horizontal analyses of accounting data.

The explanation for this phenomenon became apparent upon analyzing the accounting practices of the SMEs within the dataset. Unlike their larger, publicly listed counterparts, these smaller entities do not always adhere to the stringent accounting standards that facilitate precise financial ratio calculations. Such an informal approach to accounting often results in critical data entries being recorded as zero or values not accurately reflecting the company's operational reality, thus complicating the task of analyzing these companies through conventional financial indices.

The generation of images based on financial indices, distorted by irregular accounting practices, hampers the CNN's ability to discern meaningful patterns. In contrast, images derived from horizontal and vertical analyses tend to exhibit consistent patterns that are typical across companies of similar sizes.

This phenomenon is particularly evident in the loss metrics during the model training phase. Models trained with images created from financial indices exhibited higher loss rates compared to those trained with images from vertical and horizontal analyses of accounting data.

The most pronounced discrepancy was noted in the training of Division 47, where the model processing financial index images incurred a substantial loss rate of 24.2%. Despite Division 47 having the largest image dataset and achieving an impressive accuracy rate of 92.2%, coupled with a F1 Score of 93.2%, the high model loss signaled potential issues in the CNN training, rendering the model's reliability questionable.

Conversely, training with images based on vertical and horizontal accounting analyses yielded significantly improved results for Division 47. The model's accuracy soared to 97.5%, with the F1 Score closely following at 97.4%, and a model loss recorded at 8.5%. These metrics suggest a prediction capability that surpasses the models explored in other studies referenced in section 2.3, which focused on neural networks processing quantitative data.

Given that unreliable accounting entries can impact multiple financial ratios simultaneously, the effect on the CNN's image classification capability is substantial. However, these inaccuracies within individual accounting data points do not affect other accounts in the same way, allowing for a more reliable CNN image classification based on this data.

Consequently, the decision was made to exclusively utilize images from vertical and horizontal analyses of accounting data in subsequent training sessions.

5.3.3. Same division training and testing

The training encompassed three distinct divisions: 47, 82, and 86. It was found that the volume of images significantly influences the efficacy of the predictive model.

Divisions 47 and 86 were specifically chosen for further investigation due to their more extensive image databases, comprising 6,417 and 2,043 images, respectively. The predictive models for Division 86 demonstrated reasonable accuracy when evaluated solely based on test results. However, when considering both accuracy and the accompanying high loss rates, the models' robustness comes into question.

For Division 47, as elucidated in Section 4.2, the models constructed from images based on accounting account data yielded exceptionally reliable predictions, as evidenced by their high accuracy and test accuracy rates.

Regarding Division 82, despite employing 1,244 images, the dataset proved insufficient for the development of a model with dependable predictive power.

5.3.4. Training and testing with the intersection of two divisions

When assessed separately, Divisions 47 and 86 yielded contrasting results. Division 47 developed a high-caliber model utilizing images based on accounting account data, yet this was not replicated with images derived from financial indices, which resulted in a weaker model. For Division 86, the models were suboptimal for both sets of images.

This led to the inquiry: Could combining two disparate groups—'retail commerce' and 'human health care activities'—enhance model reliability? Accounting, a universally applicable technique, serves as a standardized method for recording financial transactions, interpretable by those versed in its principles, regardless of company type.

The created images incorporated identifiable data for both the division and group, raising the question of whether CNN could produce an improved prediction model for Division 86 when trained in conjunction with a dataset from another division with more extensive imagery.

Subsequent results indicated an augmented predictive capability for models trained on both financial data images and accounting data images. In the case of the model based on accounting account imagery, accuracy escalated to 96.4%, with F1 Score achieving 97.9% — the highest in all training iterations — and a test loss of 10.4%, thus qualifying as a robust predictive mode.

To ascertain whether the model indeed bolstered the accuracy for new data, the same set of 410 images from Division 86's solo test (which had a test accuracy of 71.5%) were employed in the model trained on the amalgamated dataset of the two divisions. This combined training yielded a significant improvement, with an F1 Score of 93,6%, thereby establishing the combined model as superior to the one trained exclusively on Division 86 data.

5.3.6. Training and testing with the intersection of three divisions

Like the prior scenarios, the stand-alone analysis of Division 82 yielded suboptimal outcomes, attributable to the limited dataset — the smallest among the three divisions examined. Nevertheless, the integrative processing of Division 82 alongside Divisions 47 and 86 demonstrated the model's proficiency in evaluating companies across disparate sectors. This integration saw a significant leap in the F1 Score, from 71.9% for Division 82 alone, to an impressive 96.3% when combined with the other divisions. This is despite a marginal loss decrease of 11% compared to the combined results of Divisions 47 and 86. The CNN, informed by the constructed imagery, proves adept at discerning the accounting attributes that underpin a company's financial well-being.

6. CONCLUSIONS

The financial market has perennially grappled with defaults and insolvencies affecting companies of all scales, creating systemic disruptions. Such uncertainties especially disadvantage small and medium-sized enterprises (SMEs) due to the obscured visibility of their financial health.

Historically, predictive analyses of defaults and bankruptcies have centered on publicly traded firms—larger entities whose data disclosure is mandated by securities regulation.

A notable limitation in existing research is the reliance on annual balance sheet data. Although the incorporation of quarterly balance sheets has advanced numerous studies, these too pertain mostly to larger corporations, leaving a data gap in relation to SMEs.

SMEs have long been relegated to a 'gray area' within the financial landscape, where scant information is a significant barrier to developing cost-effective financial products tailored for smaller enterprises. The challenge is further compounded by the difficulties these companies face in securing capital. When SMEs do access funds, the risk is often assessed superficially, resulting in higher rates than those offered to larger, publicly traded companies.

Prior research has made commendable strides using machine learning techniques, employing data on variations in financial indices. Such studies have tended to focus on changes over a specific timeframe, without delving into the progression of data. Even research incorporating quarterly data typically adopted this static approach.

The current study seeks to innovate by proposing a methodology to analyze SMEs through their accounting practices, employing neural networks with a focus on temporal data analysis. The convolutional neural network was selected for its proficiency in processing time series, offering a dynamic method to examine images representative of accounting data.

Convolutional Neural Networks (CNNs) are widely employed across various fields for their proficiency in object recognition within images. They exhibit an intrinsic capacity to learn features from data in an automatic and hierarchical manner, beginning with the extraction of fine-grained details and progressively moving towards the recognition of more complex, abstract patterns.

Incorporating images that encompass data on a company's division, group of operations, country identification, and inflation metrics—both current and over the past year—provides CNNs with a comprehensive data set to facilitate a more extensive analysis. The alignment of variables along rows in the image format explicitly represents their evolution, aiding the CNN in pattern detection. This advantage is underscored by the results shown in Training 2 and 14 of Table 7, where Training 2 used a limited dataset from a single division to achieve satisfactory accuracy, whereas Training 14 highlights the enhanced analytical power when processing images from a division with a larger dataset, even if the divisions are distinctly different.

The methodology for constructing these images marks a significant advancement for future research. It illustrates the superiority of CNNs in interpreting variations in raw variables as opposed to indices derived from such variables. CNNs have demonstrated the capability to delve deeper into unprocessed data and extract meaningful insights. This study introduces a novel approach to 'encoding' time-series data for CNN analysis, highlighting how image construction can encapsulate subtle variations in the data. This method opens new avenues for structuring information and conducting analyses in subsequent research endeavors.

While the study's findings are promising, they do acknowledge certain constraints, notably the requisite for a substantial volume of images to facilitate effective training. Despite this limitation, the CNN demonstrated its capacity to accurately classify companies, achieving an accuracy exceeding 95%.

Hence, the research offers the financial market a practical and equitable method for assessing the default and bankruptcy risk of businesses, particularly benefiting SMEs by providing a more accurate representation of their fiscal health. With a model trained across a diverse range of industries, new entities can be assessed and directly compared to an extensive trained database.

An incidental yet significant contribution of this research is the proposal of a standardization approach for accounting charts. The development of a uniform chart of accounts, coupled with the application of the Universal Sentence Encoder to detect similarities between accounting nomenclature, is an innovative strategy with broad applicability. Specifically, this methodology empowers the system to process accounting data across languages by leveraging multilingual sentence comparison.

Therefore, the approach of creating data-infused images is validated as universally applicable, with a caveat to ensure congruence in the divisions and groups when extending it to international contexts. This presents an opportunity for further research to explore multinational corporate accounting, examining the feasibility of a model trained to analyze companies across borders.